\documentstyle[epsf,aps,prb,preprint,tighten]{revtex}
\input{epsf}
\begin{document}
\draft
\title{Precursor Pairing Correlations and Pseudogaps}
\author{Mohit Randeria}

\address{Theoretical Physics Group, Tata Institute of Fundamental Research,
Mumbai 400005, India}

\address{%
\begin{minipage}[t]{6.0in}
\begin{abstract}
I begin by briefly reviewing various experimental results on
the pseudogap phenomena in underdoped cuprates.  I argue
that, taken together, all of these lead to a picture of singlet
pairing above $T_c$.  I then explore the idea that
the pseudogap is a normal state precursor of the superconducting gap
due to local, dynamic pairing correlations in a state without
long range phase coherence.  Early work on simple model systems
which exhibit pseudogap anomalies in the normal state of 2D
superconductors in a low density, small pair size regime is 
reviewed and critically re-examined in view of more recent developments.
I also describe recent studies of how
the underlying d-wave superconducting ground state
affects the anisotropy of the pseudogap and the destruction of the
Fermi surface.
\typeout{polish abstract}
\end{abstract}
\pacs{Varenna Lectures, 1997}
\end{minipage}
}
\maketitle
\narrowtext

\noindent
{\bf I. Introduction}
\medskip

The deviations from Fermi liquid theory (FLT) in the normal state of
high $T_c$ superconductors are now well accepted as experimental facts,
even if there is no theoretical consensus about their origin. 
These anomalies were first established from transport studies in 
the optimally doped materials, 
i.e., those with the highest $T_c$ within a given family. 
Angle-resolved photoemission spectroscopy (ARPES)
then showed that there is a well-defined Fermi surface even though there
are no sharp quasiparticle excitations in the normal state.
It has recently become clear that the underdoped cuprates exhibit 
even more remarkable deviations from FLT than the optimally doped 
materials: not only are the quasiparticles not defined, but even the Fermi 
surface becomes ill-defined due to the opening of a pseudogap above $T_c$. 

In these lectures, I will first describe the experimental evidence
for the pseudogap phenomena as obtained by various probes:
NMR, optics, thermodynamics, $\mu$SR, photoemission and tunneling.
I will then argue that these remarkable observations 
point to the existence of singlet pairing correlations 
above the superconducting $T_c$. I will next describe calculations
on model systems that capture some of the physics of
the pseudogap materials. The pseudogap is argued to be a natural
consequence of local, dynamic pairing correlations 
in the normal state of low density, small pair-size superconductors.

\medskip
\noindent
{\bf II. Pseudogap Experiments}
\medskip

I will briefly review experiments on the underdoped cuprates
in the pseudogap regime regime above $T_c$.
The word ``pseudogap'' is used to mean a large
suppression of low frequency spectral weight (in contrast to a 
``hard'' gap, which signifies strictly zero spectral weight).

{\bf Spin Dynamics:} Some of the earliest evidence for the 
suppression of low frequency spectral weight above $T_c$  
in underdoped materials came from NMR experiments 
[\onlinecite{NMR}]. 
The spin susceptibility $\chi(T)$,
deduced from the Knight shift, is found to be strongly
$T$-dependent, decreasing by a factor of three as the temperature
is reduced from 300K to $T_c = 60$K in underdoped YBCO, and
then decreases smoothly through $T_c$.
At the same time the relaxation
rate $1/T_1T$ on the $^{17}$O and $^{89}$Y nuclei, 
(where the form factors filter out the antiferromagnetic contribution)
was also found to be strongly $T$-dependent, with $1/T_1T \sim \chi(T)$.
Note that in conventional metals both these quantities are
$T$-independent; Pauli spin susceptibility and 
Korringa relaxation $1/T_1T \sim \chi^2$.
For the $^{65}$Cu nuclei the relaxation rate $1/T_1T$ 
shows a non-monotonic $T$-dependence. With decreasing $T$, it first increases,
presumably reflecting a build up of antiferromagnetic correlations, but 
eventually decreases below 150K, showing that a pseudogap is 
opening up even for the excitations probed by the Cu nuclei.

All of these anomalies, are collectively called ``spin gap'' behaviour. 
I emphasize that, even more so than the spin
susceptibility, it is the rapid drop in $1/T_1T$ with decreasing $T$
which directly gives evidence for a pseudogap since 
$1/T_1T = \lim_{\omega \rightarrow 0}
\sum_{\bf q}F({\bf q}){\rm Im}\chi({\bf q},\omega)/\omega$
directly probes the low frequency spectral weight in 
the spin excitation spectrum.
(Here $F$ is the form factor which is different for different nuclei). 

{\bf Charge Dynamics}: 
Early studies of the in-plane conductivity $\sigma_{\rm ab} (\omega)$ 
of YBCO as a function of doping reveal several important features 
[\onlinecite{OPTICS_AB}]. First, the
inferred scattering rate $1/\tau$ increases with underdoping, suggesting that
at high temperatures the electronic excitations in underdoped materials are 
even less sharply defined than at optimal doping. Second, there is a 
considerable suppression of $\sigma_{\rm ab}$ at finite frequencies below
500 cm$^{-1}$ which begins at temperatures much above $T_c$; the onset 
temperature for this gap-like feature increases with underdoping.
Remarkably $\sigma_{\rm ab} (\omega=250 cm^{-1};T)$ has the same 
qualitative $T$-dependence as $1/T_1T$ for $^{17}$O 
(see Fig.~4 of Rotter et al.~[\onlinecite{OPTICS_AB}]). 
Thus the charge response appears to show the same
suppression of spectral weight as the spin response. 

However, there is one crucial difference in the low frequency 
limit of the charge response. 
Recent optical data at lower frequencies show the appearance of 
a narrow ``Drude-like'' peak which shows directly that there are low lying 
charge excitations [\onlinecite{TIMUSK}], unlike in the spin channel. 
This is consistent with earlier d.c. resistivity 
studies [\onlinecite{TRANSPORT}] in underdoped systems which see
a characteristic deviation below the linear resistivity,
implying that $\sigma_{\rm ab} (\omega=0;T)$ {\it increases} slightly 
as $T$ decreases unlike the higher frequency pseudogap behavior 
noted above. We shall return later to the question of these low
lying charge excitations in the a-b plane.

There are many aspects of the c-axis transport which are puzzling, some
of which appear to be directly linked to the opening up of a pseudogap.
The c-axis optical conductivity 
[\onlinecite{TIMUSK,OPTICS_C}] 
in underdoped YBCO clearly shows loss of spectral 
weight below 500 cm$^{-1}$ whose onset temperature is much above $T_c$. 
In contrast to the in-plane conductivity, the pseudogap
$\sigma_{\rm c} (\omega;T)$ appears to directly extrapolate to
the inverse of the d.c. resistivity (i.e., there is no narrow peak at 
low frequencies). One of the puzzling features is the manner in 
which the optical conductivity sum rule is satisfied: since no weight 
is building up near $\omega = 0$ (as in the ab plane), it must be 
transferred to above the gap, but this is not apparent even up to very
high frequencies. Finally, it is found that, at fixed $\omega$ in
the pseudogap region, $\sigma_{\rm c} (\omega;T)$ tracks the $T$-dependence
of the spin susceptibility, once again showing that the spin and charge
degrees of freedom are both getting gapped out, and that the c-axis
transport is probing the pseudogap in the a-b plane excitation spectrum. 

{\bf Specific Heat:} The specific heat experiments of Loram and coworkers
on YBCO as a function of doping have been discussed in detail by 
W. Y. Liang at this school, so I will be very brief. Specific heat
probes all degrees of freedom including spin and charge excitations.
Once the background subtractions for the lattice contribution are made, 
the electronic contribution gives direct evidence for a pseudogap 
[\onlinecite{SP_HEAT}] developing above $T_c$ with
increasing underdoping. In the underdoped samples, the phase transition
looks less and less mean-field like, and furthermore,
loss of entropy begins to occur at temperatures much higher than $T_c$. 
The $T$-dependence of the entropy has been argued to be indicative
of a degenerate normal state and not consistent with the thermodynamics 
of a gas of preformed bosons (which, by definition, would be a 
non-degenerate system above its $T_c$).
A direct measurement of the electronic chemical potential 
of the underdoped systems, along the lines of [\onlinecite{VANDERMAREL}],
would be very useful in establishing the quantum degeneracy of the system.

$\mu${\bf SR:} Next we turn to the $\mu$SR measurements of the low 
temperature penetration depth $\lambda_L(0)$, which is proportional to
the $T=0$ superfluid density $n_s(0)/m^*$, by Uemura and coworkers, 
the first of which appeared as early as 1989 
[\onlinecite{UEMURA}]. 
A plot of $T_c$ versus $1/\lambda_L^2(0)$ for a large variety of materials
shows the remarkable scaling $T_c \sim n_s(0)/m^*$ for all underdoped
cuprates. This clearly suggests that transition is not of the standard
BCS mean-field variety. The connection of this important observation
by Uemura with pseudogap physics will be discussed later in Section VII.

{\bf Photoemission:}
Recent angle-resolved photoemission (ARPES) studies 
[\onlinecite{STANFORD,DING_96,DING_97}] on Bi-2212 compounds, 
have made very important contributions to our
understanding of the pseudogap phenomenon. Prior to 
this work there was no clear understanding of many issues:
Does the pseudogap only affect the spin degrees of freedom, or also 
the charge channel? (Although the optical conductivity data is quite 
unambiguous in this regard, nevertheless
it was only the spin gap which tended to be emphasized in the literature).
Does the pseudogap have any k-dependence? What is the connection, if any,
of the pseudogap, above $T_c$ with the SC gap below $T_c$?
The ARPES data provide answers to all of these questions.

The ARPES studies have found a highly anisotropic suppression
of spectral weight that persists far above $T_c$. 
Near the $(\pi,0)$ point the pseudogap is the largest
and it closes only at a much larger temperature $T^*$, which
correlates well with the crossover scale determined from other
measurements. Along the zone diagonal the pseudogap vanishes.
Remarkably, the pseudogap has the same magnitude (or scale) as the 
SC gap at each {\bf k} point (the maximum gap in optmal Bi-2212 is around
35 meV), and it evolves smoothly through $T_c$, 
with essentially the same {\bf k}-dependence just above $T_c$ 
as the SC gap below $T_c$.
The only difference in the spectroscopic data going through $T_c$.
is that there are well defined quasiparticle peaks in the SC state
but not above $T_c$. More recently it has been found [\onlinecite{DING_97}]
that both the pseudogap (above $T_c$) and the SC gap (below $T_c$)
are ``tied to'' the Fermi surface that exists above $T^*$ when all
gaps are closed. In other words there is an ``underlying'' Fermi surface 
which appears to have been gapped out in the pseudogap state. 
For a detailed review of these results, see
the companion set of lectures on ARPES  given at this school 
[\onlinecite{MR_JC}]. 
    
Very recently ARPES studies [\onlinecite{HARRIS_97}] have observed 
a pseudogap in Bi-2201, which correlates well with the SC gap in
that material (which is around 10 meV as opposed to the
35 meV scale in Bi-2212). This observation is particularly significant 
because Bi-2201 has one Cu-O layer per unit cell, and thus it shows that
bilayers are not essential to the pseudogap phenomena.

{\bf Tunneling:}
Finally we conclude this Section by mentioning some very recent STM
work by the Geneva group [\onlinecite{FISCHER}]
on Bi-2212, in which a pseudogap is observed above $T_c$ in tunneling 
spectroscopy and found to scale with SC gap. Further, the tunneling spectra
show smooth evolution through $T_c$ from the SC state to the pseudogap state.

\medskip
\noindent
{\bf III. Survey of Scenarios}
\medskip

The experiments on underdoped cuprates
described above give very clear evidence for
a SC transition that looks quite different from the BCS mean field 
transition, and also for a ``normal'' state above $T_c$, with a pseudogap,
that looks quite different from an ordinary Fermi liquid.

A common line of reasoning is that a theoretical understanding of the
non-Fermi liquid normal state must come first because only then can
one study the superconducting instability in that state. 
In these lectures, I will take a very different point of view and 
invert the question: given that the ground state is
a superconductor made up of pairs of fermions, must the corresponding
normal state necessarily be a Fermi liquid?
As I will show the answer turns out to be ``No'', and, in fact, under
very general conditions the normal state has pseudogap anomalies 
arising from pairing correlations in a state without phase coherence
[\onlinecite{NEGU_92,CROSSOVER_93,NEGU_95}].
In this scenario the pseudogap is a precursor to the SC gap. The ARPES
data, which establish the connection between the SC gap below Tc 
and the pseudogap above Tc, are a vindication of this approach.

But before turning to the main theme of these lectures, let me briefly
discuss some of the alternative approaches to understanding the pseudogap;
it is an impossible task to survey the whole literature.
One general class of approaches is what I call ``non-pairing'' scenarios
in which the pseudogap has no connection with the d-wave SC gap.
In my view, it is very unnatural for non-pairing theories 
to obtain a pseudogap scale and anisotropy which is closely tied 
to the SC gap, and also to obtain the observed smooth evolution through $T_c$.
Nevertheless, some of these ideas, particularly those related to 
antiferromagnetic correlations
[\onlinecite{KAMPF},\onlinecite{CHUBUKOV}], may be relevant to
spectral weight suppression on the energy scale of $J \sim 100$ meV 
or more (for which there is some experimental evidence in ARPES),
which is much larger than the SC gap scale (35 meV in optimal Bi-2212).

Another approach involves ``spin-charge separation'' and
the pseudogap is associated with pairing of hypothetical
$S=1/2$, charge neutral fermions called spinons, rather than
of real electrons. This approach has its origin in the early
resonating valence bond (RVB) ideas of Baskaran, Zou and Anderson
[\onlinecite{BZA}] which already contained the idea of singlet
pairing without phase coherence. It was further developed by 
Kotliar and by Fukuyama [\onlinecite{RVB1}], 
and particularly by Lee and Wen [\onlinecite{RVB2}]. 
The ``breakup'' of physical electrons into spinons and holons
(charged, spinless bosons) and their interaction with gauge fields
has been central to the more recent developments.
The hope is to understand not only the pseudogap phase,
but the entire phase diagram from Mott insulator to the overdoped regime.
It is not very clear, at least to me, if the holons and spinons
have any physical reality (in a trivial sense they do not,
since these are gauge-dependent objects) or if they are simply
a more convenient basis in which to do calculations (actually
the problem turns out to be strongly coupled even in this basis).

It thus seems useful to ask if the intuitively appealing, original
RVB picture of phase-incoherent singlet pairs can be studied in a 
simpler setting, with the more modest goal of understanding the
pseudogap regime, rather than the full phase diagram. It is in this
sense that the results discussed below tie in with the RVB idea.

I close this Section by pointing out some other proposals that
have something in common with the pairing correlation ideas
to be described next.  The work of Emery and Kivelson
[\onlinecite{EMERY}] on classical phase fluctuations determining
the underdoped $T_c$, and its relation to the present approach,
will be discussed in detail in Section VII. 
In an early paper Doniach and Inui discussed the possibility of
the insulator going into a state with incoherent pairs upon doping
[\onlinecite{DONIACH}]. There are also several proposals
involving mixed boson-fermion models, some of which
[\onlinecite{RANNINGER}] will be discussed elsewhere in
the proceedings of this School, and others [\onlinecite{GIL}]
will be mentioned later in Section VIII.

\medskip
\noindent
{\bf IV. Pairing Correlations above $T_c$}
\medskip

One of the reasons why the pseudogap experiments described in
Sec.~II seem surprising to us is that we are conditioned
to think about superconductors in terms of the BCS mean field theory.
There are two aspects to the BCS description: (a) pairing, which leads to
a gap in the spectrum, and (b) phase coherence, or 
macroscopic occupation of the ${\bf q} = 0$ pair state, which leads to 
the superflow properties. It is a very special feature of BCS theory,
and of the conventional metallic superconductors so well described by
this theory, that both of these effects happen together at the same
temperature $T_c$: as soon as the pairs form they automatically
condense. One only needs to look at other phase transitions in nature,
say metallic magnetism, to see that this is an exception rather than
the rule. In magnets the formation of the moments, the analogue of (a),
and their ordering, the analogue of (b), occur at widely separated
temperature scales: the Stoner temperature and the Curie temperature
respectively.

The small parameter that justifies the BCS mean field approximation 
for conventional superconductors is $T_c/E_f \ll 1$ or $1/k_F\xi_0 \ll 1$:
each pair contains a large number of other pairs within it. 
Given the low carrier concentration, and the small pair size in the 
cuprates, it is hardly surprising that one has to re-examine the validity
of this approximation. In order to go beyond BCS theory, into a regime
where there is no small parameter it is useful to consider the problem
of the crossover from BCS pairing to Bose-Einstein condensation of
tightly bound pairs.  I will be very brief on this topic and refer
the interested reader to ref.~[\onlinecite{CROSSOVER_REVIEW}]. 
I emphasize that I am not claiming that by varying any experimental
parameter in the high $T_c$ cuprates one is traversing such a crossover.
The crossover picture allows us to
bracket an interesting intermediate coupling problem ($\xi_0 \sim k_F$)
between two known, and rather different, limiting cases (BCS and Bosons).

The simplest lattice model within which this problem can be studied
is the attractive Hubbard model where the pair-size can be tuned by
varying the strength of the on-site attraction $|U|$ relative
to the nearest neighbour hopping $t$:
\begin{equation}
\label{eq:NegU}
H = -t\sum_{i,j;\sigma} c_{i\sigma}^\dagger c_{j\sigma} + {\rm h.c.}
- |U|\sum_{i} n_{i\uparrow}n_{i\downarrow} 
+ \mu \sum_{i;\sigma}n_{i\sigma}.
\end{equation}
The chemical potential $\mu$ is used to fix the average 
density $\langle n \rangle$. 

In Fig.~1 we show a schematic phase diagram of this model 
(away from half-filling so that the ground state is a superconductor). 
The weak coupling $|U|/t \ll 1$ limit is very well described by BCS theory.
The pair size $\xi_0 \gg a$ (the lattice spacing) and upon increasing 
$T$ the destruction of pairs and of phase coherence occurs at the same 
$T_c \sim t\exp (-t/|U|)$. The normal state is a Fermi liquid.
In the opposite extreme where $|U|/t \gg 1$ the ground state is
a condensate of hard core lattice bosons. Upon heating,
phase coherence is lost at $T_c\sim t^2/|U|$ (determined by the
effective hopping for the bosons) and the state
just above $T_c$ is a normal Bose liquid of tightly bound pairs.
It is only at a much higher temperature $T^\ast\sim |U|$ that
the pairs dissociate. Our main interest here will be the intermediate
coupling regime, particularly above $T_c$.

A closely related continuum model has been studied in some detail
in refs.~[\onlinecite{CROSSOVER_93,CROSSOVER_97}] where a functional
approach was used to study the global phase diagram at all values
of the temperature and coupling. It was found that a mean field 
approximation is able to capture the physics of the SC ground state,
and excitations, including collective modes, all the way from the
BCS to the composite boson limit at $T=0$ [\onlinecite{CROSSOVER_97}].
However, such an approximation is completely inadequate to 
describe even the qualitative physics of the normal state 
as one moves away from the weak-coupling BCS limit. 
There is a crossover temperature $T^*$ above 
which the system can be described in terms of free fermions, but below which
pairing correlations become important, and a simple
mean field description fails. With increasing coupling, or decreasing
pair-size, $T^*$ and the transition temperature $T_c$ separate from each
other and a proper treatment of fluctuations about the
trivial saddle point is necessary to understand the temperature
region between $T^*$ and $T_c$ [\onlinecite{CROSSOVER_93,CROSSOVER_REVIEW}].
For sufficiently strong coupling, in {\it any} dimension,
$T^* \gg T_c$ and it is dynamic (frequency dependent) fluctuations
which lead to this separation of scales. In 2D this separation
between scales persists to much weaker coupling for two reasons:
classical phase fluctuations, and the fact that arbitrarily weak
attraction leads to the formation of independent bound states
[\onlinecite{RDS1}].

The functional integral approach described above led to a very 
useful interpolation scheme that gave reliable answers 
in both the BCS and Bose limits. However,
it was least reliable in the intermediate
coupling normal state where there is no small parameter to control
the calculation. This, however, is the most interesting regime, since
it is here that we expect to see characteristic deviations from Fermi
liquid theory (FLT) as it crosses over from a weak coupling Fermi liquid
limit to a strong coupling normal Bose liquid limit.
The question then is whether there is a broad intermediate
coupling regime in between these two limiting
cases, especially in 2D, where the normal state is
a degenerate Fermi system and yet exhibits non-Fermi liquid correlations
for $T_c < T < T^*$? And, if so, what are the characteristic
deviations from FLT due to pairing correlations? 
We will use quantum Monte Carlo simulations
to address this question in the next Section.

\medskip
\noindent
{\bf V. Pseudogap in s-wave Models} 
\medskip

I now describe results on the anomalous normal state
properties of the 2D attractive Hubbard model which have been
obtained from extensive quantum Monte Carlo (QMC)
simulations [\onlinecite{NEGU_92,NEGU_95,NEGU_95B}]. 
The finite temperature determinantal QMC, developed
by Scalapino and co-workers 
[~\onlinecite{SCALAPINO}], begins by writing down the
coherent-state path integral for the lattice Fermi system with the
imaginary time direction discretized. The (quartic) interaction term is 
decoupled using a discrete Hubbard-Stratonovich decoupling, using
Ising variables as the bosonic fields [\onlinecite{HIRSCH}]. The Fermi fields
are then integrated out leading to an effective bosonic field theory
on a discrete space-(imaginary) time lattice. The functional integral
over the bosonic (Ising) fields is then performed using Monte Carlo
techniques, and various fermionic correlation functions are calculated. 

This QMC method is uniquely suited to the problem at hand
for a variety of reasons.  First, it is an inherently non-perturbative, 
finite temperature technique.  Second, it works best in the intermediate 
coupling regime (in weak coupling, the pair-size is larger than the 
finite size systems one can study, and in strong coupling there are 
large Trotter errors introduced by the imaginary time discretization 
and also the very slow exploration of phase space). Finally, the attractive
Hubbard model does not have a ``sign problem'' at any filling,
unlike most other interesting fermionic systems, thus making it possible to
obtain very accurate results. This comes about due to
a decoupling scheme introduced by Hirsch [\onlinecite{HIRSCH}] such that the
weight factors in the resulting bosonic theory are positive definite.
There are, however, two limitations of all QMC calculations that
we must face up to: finite system size and imaginary time.
We have studied the properties of interest on lattices 
of typical size $8 \times 8$, and in many cases up to $16 \times 16$,
to ensure that our results are not finite size artifacts. 
The size limitation is not very severe because
we are interested in normal state physics, rather than that
of the phase transition (at which the correlation length diverges).
As regards analytic continuation from imaginary time data to real 
frequencies, this is a real limitation. In our own work we have developed
some tricks for looking at some correlation functions, like the
density of states at the chemical potential, and the NMR relaxation
rate that allows us to finesse this problem under certain
circumstances; see [\onlinecite{NEGU_92,NEGU_95}]. But there are no useful
results available on many interesting quantities such as the d.c. resistivity.
More recently there have been some results on spectral functions and on
density of states obtained by using the maximum entropy method to
do the analytic continuation [\onlinecite{SINGER}].

The normal state anomalies described below are
in fact seen over a broad range of parameters. Here I focus on results 
at a density $\langle n \rangle=0.5$, at a moderate coupling strength of 
$|U|=4t$. For this choice, we have the rough estimate $k_F\xi_0 \sim 3$.
All statistical error bars not explicitly shown in the 
figures are of the size of the symbols.
From the decay of the SC order parameter correlation function 
we estimate $T_c \simeq 0.1t$; from the spin susceptibility we will
deduce below that $T^\ast \sim 1t$. 

The uniform, static spin susceptibility $\chi_s$
as a function of $T$ is shown in Fig.~2. In a Fermi liquid
$\chi_s$ should be $T$-independent
(Pauli susceptibility) at temperatures smaller than the Fermi energy.
In contrast, we find a strongly $T$-dependent
result with $d\chi_s/dT > 0$ below a temperature scale $T^\ast \simeq 1t$
(in fact, we may use this to define the crossover temperature $T^\ast$). 
The tendency toward strong singlet pair fluctuations in the normal state
leads to $d\chi_s/dT > 0$. This behavior is easy to understand at
 large $|U|$ where tightly bound singlet pairs do not contribute
to $\chi_s$ unless they are ionized. What is remarkable is that
such correlations persist down to $|U| = 4t$
where one has a degenerate Fermi system (as shown below).

I next turn to the single-particle density of states (DOS)
$N(\omega)$ where $\omega$ is measured from $\mu$. To obtain
this quantity from QMC requires analytic continuation which is
usually a highly nontrivial problem. We have devised a simple
method [\onlinecite{NEGU_95}] to estimate 
$N(0)$ for temperatures much smaller than 
the frequency scale on which $N(\omega)$ shows significant variation.
Thus in the absence of a ``low frequency scale'' we are able to
extract $N(0)$ from the local Greens function at imaginary
time $\tau = \beta/2$. The resulting $N(0)$ is plotted in
Fig.~2 as a function of temperature.
We see that with decreasing $T$ a pseudo-gap develops
at the chemical potential in the normal state, i.e., the DOS is depleted.
To emphasize this $T$-dependence of the DOS we use the notation $N_T(0)$.
It is interesting to note that $\chi_s(T) = N_T(0)$ (to within
the errors inherent in extracting the latter). While this equality
looks like that in Fermi liquid theory (FLT), 
note that both quantities are strongly $T$-dependent, 
in marked contrast to FLT. 

We have studied the NMR relaxation rate $1/T_1$
(using an analytic continuation method similar in spirit to
the one described above for the DOS) and found [\onlinecite{NEGU_92}]
that, unlike the Korringa behavior expected of usual metals, 
$1/T_1T$ is strongly $T$-dependent
and $1/T_1T \sim \chi_s(T)$ as seen in Fig.~3.
This behavior of $1/T_1T$ is characteristic of a spin gap
as discussed in connection with the experiments in Sec.~II.
Since there is no non-trivial momentum space structure 
in this model, we cannot discuss the differences between 
relaxation rates of various nuclei, which depends on a proper
description of short range antiferromagnetic correlations.
Our discussion only sheds light on those aspects of the data
which are not affected by these AFM correlations.

To understand the origin of
$1/T_1T \sim \chi_s(T)$ we study
$K({\bf q}) = \lim_{\omega \rightarrow 0}{\rm Im}\chi({\bf q},\omega)/\omega$.
We find [\onlinecite{NEGU_95}] that $K({\bf q})$ is of the form
$N_T(0)/\Gamma({\bf q})$ where $\Gamma \sim v_Fq$ is
essentially the same as that for the noninteracting system. Thus
$K({\bf q})$ is suppressed more or less uniformly in ${\bf q}$-space
with decreasing temperature and its $T$-dependence is governed by the
pseudo-gap in the DOS. Using  $1/T_1T = \sum_{\bf q}K({\bf q})$,
the above form of $K$ gives a natural explanation of
$1/T_1T \sim \chi_s(T)$.

Does the charge channel also exhibit this pseudo-gap? To answer this
question we analyze the compressibility $dn/d\mu$ obtained
by numerically differentiating
the average density $\langle n \rangle$ measured as a function of $\mu$.
(We found that $dn/d\mu$ shows much larger finite size 
effects than, e.g., $\chi_s$, which forced us to work
on large lattices ($L = 16$)).
We see from Fig.~4 that, in sharp contrast to the one-particle DOS,
$dn/d\mu$ is very weakly $T$-dependent, i.e., there is no pseudo gap in 
the charge channel. While this $T$-independence superficially looks like
FLT, we argue below that in fact it arises due to collective excitations.

It is perhaps worth noting that a simple RPA analysis (particle-hole bubbles) 
fails to account for the observed behavior in either the spin
or charge channels (except for the general trend that 
attractive interactions lead to a decrease in $\chi_s$ and an increase 
in $dn/d\mu$). For the susceptibility
$\chi_{\rm rpa} = N^0/(1+|U|N^0)$, where $N^0(0)$ is the noninteracting 
DOS, RPA fails to account for the $T$ dependence. On the other hand
$dn/d\mu^{\rm rpa} = 2N^0(0)/\left[1 - |U|N^0(0)\right]$ 
predicts an entirely spurious instability (phase separation) at $|U|N_0(0) = 1$.
In fact, pairs form at large $|U|$ and their residual interactions
are repulsive, so that the compressibility of the system remains finite
for {\it all} $|U|$. We have calculated [\onlinecite{NEGU_95}] 
the compressibility at $T=0$ within a
mean field (Hartree-Fock-Bogoliubov) approach which is capable
of dealing with the $T=0$ BCS-boson crossover. We find that it has
the same order-of-magnitude as the normal state MC result
(note that we do not expect $dn/d\mu$ to
change dramatically as $T$ goes through $T_c$).
This lends further support to the argument that collective excitations
(singlet pair fluctuations) contribute to the charge channel.

Note that we find qualitative differences between the spin and 
charge response functions.  Quantitative differences are well
known even in a Landau FL where $\chi_s$ and $dn/d\mu$ are
renormalized by different FL parameters.
What we see here is much stronger: as a result of strong interactions,
$\chi_s$ and $dn/d\mu$ acquire {\it qualitatively different} $T$-dependences.
 The spin response is dominated by 
single-particle excitations ($S=1/2,Q=1$); this is $T$-dependent because
the DOS for these excitations is depleted by collective excitations, 
i.e. pairs with $S=0,Q=2$. The collective excitations, on the
other hand, directly contribute to the charge channel.
At this point it may be useful to
recall the discussion of the transport experiments in Sec.~II,
where clear evidence for low lying charge excitations was found,
in spite of the fact that the single-particle excitations were
(pseudo)gapped. 

The momentum distribution function $n({\bf k})$ is shown in Fig.~5.
Note the rapid variation with ${\bf k}$,
the remnant of a ``Fermi surface'' though it is broadened by the
temperature and by interactions through the pseudogap effect.
This, together with the fact that the chemical potential $\mu \gg T$ (see
ref.~[\onlinecite{NEGU_92}]), 
clearly shows that we are dealing with a highly degenerate
Fermi system (and not with a system of preformed bosons!). Thus it
makes sense to talk of deviations from Fermi liquid behaviour when
discussing the above results.
We thus see that the correlations in
this regime show very interesting behavior which is intermediate between
a Fermi liquid and a Bose liquid.

The DOS pseudogap in the normal state of the attractive
Hubbard model has also been verified by self-consistent
$T$-matrix calculations [\onlinecite{MICNAS}] and by analytic
continuation of Monte Carlo data using maximum entropy 
methods [\onlinecite{SINGER}]. The latter study also sees evidence for
the precursor of the Bogoliubov-like dispersion of excitations
associated with the pseudogap above $T_c$. There is an
interesting recent paper [\onlinecite{OLEG}] which introduces
analytically tractable approximations leading to a DOS pseudogap.


\medskip
\noindent
{\bf VI. Commonly Asked Questions:}  
\medskip

Before proceeding further, it may be useful to pause and answer
several questions which are commonly asked of the author. 

(1) {\bf Does the pseudogap arise from ``superconducting fluctuations''?}
The answer is no, at least not in the conventional sense that the phrase
``SC fluctuations'' has been used in the literature. The conventional
usage follows Aslamazov and Larkin [\onlinecite{LARKIN}]
where it is used to describe small, but singular, corrections to weak
coupling BCS theory in the vicinity of $T_c$. The choice of diagrams
(the four shown in Fig.~1 of [\onlinecite{LARKIN}]) is controlled by a
leading order perturbation theory in
the small parameter $\max(T_c/E_f, 1/E_f\tau) \ll 1$, and these lead to
significant effects in a small range of temperatures $(T-T_c)/T_c \ll 1$,
where the SC correlation length is diverging.

In contrast, the pseudogap anomalies are {\it not} small corrections to
a normal state background; these are large effects over a large temperature
range (persisting up to temperatures which are several times $T_c$) 
and thus lie beyond the standard perturbative SC fluctuation approach.
Nevertheless, the standard approach does contain within it the first hint 
of the pseudogap anomalies, as shown by Varlamov and coworkers 
[\onlinecite{VARLAMOV}], who have emphasized the importance 
of the ``DOS'' diagram:
where the SC fluctuation propagator acts as a self energy dressing the 
electron propagator. A very similar approach has been taken by Rainer
and coworkers (in largely unpublished work) who study systematic 
perturbative corrections to Fermi liquid theory [\onlinecite{RAINER}].

(2) {\bf Is two dimensionality and Kosterlitz-Thouless physics sufficient
to give rise to the pseudogap?}
The answer is no, by itself it is not sufficient. To see this most
clearly consider a weak coupling BCS superconductor. Let us
denote the BCS mean field transition temperature by $T_c^0$, 
and the superfluid stiffness $D_s(T=0) \simeq \hbar^2 n_s(0)/m^*$. 
In the weak coupling limit $T_c^0/E_f \ll 1$ 
which means that $D_s(0) \gg T_c^0$.
We use the universal jump condition of Kosterlitz-Thouless (KT) theory
$\hbar^2 n_s(T_c^-)/m^* = 2T_c/\pi$ to determine the actual (KT) $T_c$.
To make a rough estimate, it suffices to use the mean field form 
$D_s(T) \simeq D_s(0)(T_c^0 - T)/T_c^0$ in this condition. 
We thus find that $T_c = T_c^0 [1 - {\cal O}(T_c^0/E_f)]$; (a more
accurate theory gives a multiplicative logarithm in the correction term
[\onlinecite{Tc2D}]). Since the KT $T_c$ and the mean field $T_c^0$ are
so close to each other, there is no significant temperature range over which
pseudogap anomalies can exist in the weak coupling limit even in 2D.
This clearly points to the importance of being in a regime outside
the weak coupling BCS limit.

(3) {\bf Does the pseudogap arise from ``preformed bosons''?}
Again the answer is no, insofar as the experiments or
the calculations presented above are concerned. 
Certainly if all electrons were tightly bound into
``preformed boson'' pairs, this would lead to a gap in the
one-electron spectrum. But in this case the normal state of the system
would be a completely non-degenerate system (since $T_c$ in Bose systems
corresponds to the onset of degeneracy: thermal deBroglie wavelength
becoming comparable to the interparticle spacing).  To the extent that
thermodynamic and spectroscopic studies, described above, 
find the underdoped cuprates to be (consistent with) 
a degenerate Fermi systems, a bosonic description is untenable. 
Another important point to mention here is one made by Leggett long ago,
and described in some detail in [\onlinecite{RDS}]: in a large class of models
the observation of nodes (zeroes) of the SC gap is by itself sufficient 
to rule out the ``bosonic'' regime.

The upshot of all this is that one is neither in the BCS limit
where small deviations from weak coupling theory can be understood in
terms of the conventional theory SC fluctuations, nor is one in
a preformed boson limit, where the state above $T_c$ is a normal
Bose liquid, but rather in an intermediate coupling regime.

(4) {\bf Why is a simple model like the attractive Hubbard model
relevant for the problem at hand?} The first point to make is that
the 2D attractive Hubbard model is {\it not} a microscopic model for
the high $T_c$ cuprates; this cannot be overemphasized.
It just happens to be the simplest lattice model in which:
(a) one can study an intermediate coupling superconductor with
$\xi_0 \sim k_F$; (b) one can see a clear separation between
$T^*$, the crossover scale below which effects of pairing correlations 
become significant, and the transition temperature $T_c$ at which
phase coherence is established; and (c) one can reliably calculate 
at least some experimentally interesting properties and
establish pseudogap anomalies in the temperature range $T_c < T < T^*$.

The simplicity of this model certainly helps in that a new 
physical situation is being studied, and also it
focuses attention on the important ingredients:
low density, short coherence length, low dimensionality.
Any other more realistic model which has a pseudogap arising
from pairing correlations above $T_c$ will have to analyzed in
an analogous manner. 

At the same time, we must also remember the limitations of
a simple model. An obvious one is that the attractive Hubbard model 
has s-wave pairing. The study of similar models but with a 
d-wave ground state is comparatively recent, and will be discussed
in Section VIII.  Another simplifying feature is that we are not
capturing some aspects of the strong correlation problem.
As a concrete example, note that in the simple models considered 
here, the electronic density $n$ determines both the size of the
underlying Fermi sea, $k_F$, as well as the zero temperature
superfluid stiffness $n_s(0)/m^*$ (the importance of which will be
made clear in next Section). In the underdoped cuprates,
which are doped Mott insulators with $x$ doped holes per Cu,
the size of the Fermi sea may be large, scaling like $(1+x)$,
(satisfying Luttinger's theorem), and yet the superfluid density
may be small $n_s(0)/m^* \sim x$. 

\medskip
\noindent
{\bf VII. Phase Fluctuations and} $T_c$
\medskip

In all the calculations reviewed above we looked at the properties of the
system {\it coming down in temperature} from above $T^*$ to below it,
and asked how pairing correlations without long range order (LRO) affect
normal state properties. It is very useful to ask the complementary
question: what are the excitations about the SC ground state that
destroy the SC LRO as the {\it temperature is increased}, and thus 
determine $T_c$? 

An answer to this question was given recently by 
Emery and Kivelson (EK) in a very nice paper [\onlinecite{EMERY}].  
I will first describe their argument, 
clarify its relationship to the calculations described in Section V,
and finally discuss its relevance to d-wave systems.
Recall that in BCS mean field theory, the excitations responsible for loss
of order are the (Bogoliubov) quasiparticles coming from thermally
broken pairs, which lead to a collapse of the self-consistent gap at $T_c^0$.
EK argued, in a model-independent manner, that in a system with small
superfluid stiffness $n_s(T=0)/m^*$ phase fluctuations
are soft and define a phase disordering temperature. For a layered system
this is given by $T_c^{\theta} \sim \hbar^2 n_s(T=0)d/m^*$, where
$d$ is the layer spacing. (This is the analogue of the statement that
in a magnetic system with well defined moments, the disordering temperature
scale is set by the exchange coupling $J$). Further, the actual $T_c$
is given by the smaller of $T_c^{\theta}$ and $T_c^0$. In
the underdoped materials, which are lightly doped Mott insulators
with a low superfluid stiffness, $T_c = T_c^{\theta} \sim n_s(T=0)/m^*$.
This also accounts for the Uemura scaling [\onlinecite{UEMURA}]. 

EK go on to emphasize that the coherence length or
pair size $\xi_0$ does not enter their estimate, and hence their
result describes a completely different physical situation from that
discussed in Section V. I believe that this is not correct, for their
calculation is based on the implicit assumption that the gap is the largest
energy scale in the problem: $\Delta_0 > T_c^{\theta}$ (or else BCS
gap collapse would determine the phase transition). Writing 
$\Delta_0 \sim \hbar v_F/\xi_0$ and $n_s(T=0) \sim k_F^2/d$,
the above inequality yields $k_F \xi_0 < 1$, which is {\it the same
regime as the one we have been discussing}. Thus by working in the
pair size comparable to interparticle spacing regime in simple
model systems, we were simulating one of the key relevant conditions 
in a lightly doped Mott insulator where the superfluid stiffness is much
smaller than the SC energy gap. (See also remarks at the end of Section VI).

For a system with an isotropic s-wave gap, phase fluctuations
(in the sense of excitations of an XY model)
are clearly the most important excitations {\it not} included in a simple
mean field description. However, in a d-wave system with nodes in the gap
the situation is even more interesting. As first pointed out by 
Lee and Wen [\onlinecite{LEE_WEN}]
in a phenomenological analysis (which is quite independent of their
microscopic ideas on spin-charge separation [\onlinecite{RVB2}]), 
the SC LRO of the ground state is
most effectively lost due to the excitations around the point nodes.
These excitations lead to the well-known linear $T$ reduction in the 
superfluid stiffness (this much is true even for a d-wave BCS SC).
But since $D_s(T=0)$ is small to begin with (and this is the non-BCS part
of the argument), this linear decrease by itself is sufficient to drive the 
system normal, without affecting the gap! 
This appears to lead to an even lower $T_c$ estimate than that obtained
from phase fluctuations alone, and it too naturally leads to 
the Uemura scaling.

Whether the $T$-dependence of the $n_s(T)$ is dominated
by phase fluctuations themselves or by excitations near the nodes,
the main conclusion is that, in the regime of interest, it is the
vanishing of the superfluid density that controls the transition
and determines $T_c$, and {\it not} the collapse of the gap (which is the
main driving force in BCS). The normal state above $T_c$ must then
still have a remnant of the superconducting gap, which is precisely the
pseudogap scenario discussed in Section V. I next turn to the 
very important issue of how those results are modified when
the ground state is a d-wave superconductor.


\medskip
\noindent
{\bf VIII. Pseudogaps in d-wave models}
\medskip

At this point it would be appropriate to discuss the d-wave analogue of the
results described in Section V. However, there is no simple known model
with a d-wave SC ground state on which reliable quantum Monte Carlo
calculations can be done. (The main problem is the fermion
sign problem in quantum Monte Carlo simulations). Thus a different approach
is required. In an ongoing collaboration
with Engelbrecht, Nazarenko and Dagotto, we have used self-consistent,
conserving approximations. All of the results described below are based on
ref.~[\onlinecite{JAN_97}].
Let us consider a simple tight binding model on a 2D square lattice
with dispersion $\epsilon_{\bf{k}}=-2t(\cos k_x + \cos k_y)$, ($t=1$), 
described by
\begin{equation}
\label{eq:H}
H\!=\!\sum_{{\bf{k}},\sigma}(\epsilon_{\bf{k}}\!-\!\mu)
c^{\dagger}_{{\bf{k}}\sigma} c_{{\bf{k}}\sigma}
\!+\!\frac{1}{N}\!\sum_{{\bf k}{\bf k}'{\bf q}}
\! V_{{\bf k},{\bf k}'}
c^{\dagger}_{{\bf k} \uparrow}
c^{\dagger}_{{\bf q\!-\!k} \downarrow}
c_{{\bf q\!-\!k}' \downarrow}
c_{{\bf k}' \uparrow}
\end{equation}
where the separable potential
$V_{{\bf k},{\bf k}'} = 
-|U_d| (\cos k_x - \cos k_y)(\cos k_x' - \cos k_y')$,
is chosen to produce a d-wave SC ground state. The goal is
to study the normal state properties of this model in a regime
where $|U_d|$ is comparable to the bandwidth, and the system
is presumably dominated by pairing correlations.

We use the self-consistent $T$-matrix method 
in which we study the renormalization of the Green's functions
by the exchange of pairing fluctuations. (This method has been used 
earlier by several authors to study the normal state of s-wave 
superconductors [\onlinecite{MICNAS,TMATRIX}]).
This requires a numerical solution of the following set 
of integral equations.
The p-p channel vertex $\Gamma$ is given (symbolically) by
\begin{equation}
\Gamma = I - I G G \Gamma
\end{equation}
where $I$ is the p-p irreducible vertex and $G$ is the fully
renormalized Green's function 
\begin{equation}
G = [G_0^{-1} - \Sigma]^{-1} 
\end{equation}
defined in terms of the self-energy
which satisfies
\begin{equation}
\Sigma = V G G G \Gamma
\end{equation}
where $V$ is the bare two-body potential.
These three equations are formally exact, and to make
progress we need to specify an approximation for $I$;
we make the simplest choice $I \simeq V$.
In addition to being fully self-consistent, this
approximation is also conserving in the sense of Baym-Kadanoff.

It is important to remember, however, that there is no small parameter
which controls this calculation, and there is an on-going
debate in the literature about the ``best'' set of diagrams to
retain [\onlinecite{VILK}]. This can really only be settled by 
comparing one's favourite approximation with a more exact calculation 
such as quantum Monte Carlo, if it exists.
Even though the details of the results may well
depend upon the precise approximation scheme, I will emphasize those
qualitative aspects which I believe to be robust on physical grounds.
Also, some of the issues that arise in interpreting the
results are clearly much more general, and go beyond
the specific model or approximations being discussed here.

The main quantity of interest is the
one-electron Greens function and the associated spectral function 
$A({\bf k},\omega) = -(1/\pi) {\rm Im}G({\bf k},\omega+i0^+)$, 
which is closely related to the ARPES intensity [\onlinecite{MR_95}].
(I will not discuss here results on response functions like
the spin susceptibility, for lack of space).
From the spectral function we will deduce interesting results about pseudogaps
and about the ``Fermi surface''.  
The question of defining a ``Fermi surface'' at finite temperatures,
especially in the absence of sharp quasiparticles, is one of general interest
in the field of high $T_c$ superconductors.  (See the lectures on ARPES
by the author and Campuzano at this summer school for more discussion 
on this issue [\onlinecite{MR_JC}]).
We may identify points ${\bf k}^*$ in the Brillouin zone
at which the spectral functions 
disperse through the chemical potential ($\omega = 0$), i.e.,
$A({\bf k}^*,\omega)$ has a dominant peak at $\omega=0$.
The ``locus of gapless excitations'' $\{{\bf k}^*\}$ then 
generalizes the notion of a ``Fermi surface'' (FS). 
For a strongly correlated system at finite temperatures,
this locus need not posses all of the properties that we associate
with the usual $T=0$ definition of a FS in a Fermi liquid.
For instance, there is no guarantee that its shape or the volume
enclosed by it is $T$-independent, and we indeed find this in our numerics.
One may question this whole concept, but experimentally [\onlinecite{MR_JC}]
it has been found in the optimally
doped cuprates that a similar definition yields a
``Fermi surface'' which has many nice features. It is $T$-independent,
it forms a continuous contour in the repeated zone scheme which is quite
similar to what band theory predicts, and there is a rapid variation
of the momentum distribution as one crosses this locus.

I will now discuss the results for $U_d=-8$ and $n=0.5$
(quarter-filling) shown in Figs.~6 -- 8; qualitatively similar results 
were found for other parameter sets. 
For details of the calculations, see ref.~[\onlinecite{JAN_97}].
All the results are in the non-superconducting state above $T_c$ 
as ascertained from the pair susceptibility, and in a degenerate
regime, as seen from $\mu \gg T$ and studies of the momentum
distribution $n({\bf k})$.

The form of the interaction $V_{{\bf k},{\bf k}'}$ is such that electronic
states near the zone diagonal $(0,0) \rightarrow (\pi,\pi)$ are unaffected
by the interaction, while those near the zone corners $(\pi,0)$ are very
strongly affected. In fact, we find that near $(\pi,0)$ the spectral
functions acquire very large widths and a quasi-particle description
is not valid at any temperature. Nevertheless, at high temperatures,
there is an identifiable dispersion of the broad $A({\bf k},\omega)$ peaks
(see Fig.~6)
from which we can determine a ``Fermi surface'' $\{{\bf k}^*\}$,
even in the absence of well-defined quasiparticles, as discussed above. 
It forms a continuous contour in the
repeated zone scheme, although, unlike the experiments [\onlinecite{MR_JC}],
it is not $T$-independent in our calculations.

As the temperature is lowered below a scale $T^{*} \simeq 1$
there is a qualitative change in the spectral functions; see Fig.~7.
First the lineshape: the spectral weight of the one 
broad feature at high $T$ is now redistributed into a multiple 
peak structure. We can get further insight into the lineshape
by studying the real and imaginary parts of the
self-energy $\Sigma$ as functions of $\omega$; for details
see [\onlinecite{JAN_97}].
Second, the dominant peaks of $A({\bf k},\omega)$ now exhibit 
very anomalous dispersion. As ${\bf k}$ varies from $(0,0)$ to $(\pi,0)$, 
the spectral peak approaches $\omega=0$
but never crosses it, either ``bouncing back'' towards negative $\omega$
or ``stalling'' (depending on parameter values).

This is a clear signature of a Fermi surface crossing being destroyed
due to the opening up of a gap, even though there is no long-range
order (more appropriately, algebraic order for the 2D model under study).
This is very similar to the pseudogap seen by ARPES experiments 
on underdoped cuprates in the temperature regime $T_c < T < T^*$. 
In the spirit of the ARPES experiments
we estimate the pseudogap magnitude by making scans through
${\bf k}$-space and noting the position of the spectral function
peak which is farthest to the right, i.e. at the largest frequency
below zero. We then plot the spectral function pseudogap
$\Delta_{ps}$ as a function of $\theta=\arctan(k_y/k_x)$
in Fig.~8 at two temperatures $T=0.2t$ and $T=0.75t$.

The first point to note in Fig.~8 is the strong anisotropy of the pseudogap,
which is suppressed to zero in an arc about the diagonal.
The extent of the nodal arc region, and the magnitude
of the maximum gap at low $T$ are both sensitive functions of the choice
of parameters as seen in the inset to this Figure.
The second important point to note is the $T$-dependence of the pseudogap
anisotropy which shows that small gaps are destroyed at lower 
temperatures compared with larger gaps. This leads to Fermi arcs (of nodes)
which expand with $T$, as seen from Fig.~8. 
At sufficiently high temperatures, when the ordinary 
dispersion of the spectral functions has been restored, these disconnected 
arcs must merge to form the continuous Fermi contour above $T^*$.

All of these effects are rather remarkable. First consider the
$T$-dependent anisotropy of the pseudogap. Contrast this with gap
anisotropy in the broken symmetry state of a d-wave BCS superconductor, 
where the self-consistent gap equation leads to the solution
$\Delta_{\bf k}(T) = \Delta_0(T)(\cos k_x - \cos k_y)$. Thus the temperature
and ${\bf k}$ dependences factorize, and there is no variation of
the anisotropy with $T$; gaps at different ${\bf k}$'s vanish at the
same temperature $T_c$ defined by $\Delta_0(T_c) = 0$.
The behaviour of the pseudogap is qualitatively
different in that small gaps collapse at lower temperature.

Viewed as a function of decreasing temperature, one obtains the picture
of a continuous Fermi surface contour being destroyed in patches
as the pseudogap begins to develop starting at the $(\pi,0)$ point
and then proceeding towards the diagonal directions.
This qualitative prediction of our calculations 
has been verified in very recent ARPES experiments by the
Argonne-UIC group [\onlinecite{NORMAN}]. A schematic representation of the
the Fermi Surface destroyed in patches is shown in Fig.~9.
Although the novel $T$-dependence discussed above was first
described in [\onlinecite{JAN_97}], the general picture 
shares some qualitative similarities with a
phenomenological model proposed in ref.~[\onlinecite{GIL}], which consists
of bosonic pairs around $(\pi, 0)$ and fermions in the rest
of the zone interacting via a ${\bf k}$-dependent coupling.


\medskip
\noindent
{\bf IX. Open Questions}
\medskip

There are many open questions even within the precursor pairing
picture of the pseudogap. I will briefly list them here.
The reason why the answers are not known, even for the simple 
models considered, is the lack of reliable, controlled calculations in the
parameter regime of interest. Thus one is faced with one of the main
dilemmas of the whole field of high $T_c$ theory. On the one
hand, one needs to simplify models so that they become more
tractable, on the other hand, one probably needs to incorporate
more realistic features of the materials in order to better confront 
experiments.

Transport in the pseudogap regime is a very important open problem
for the pairing scenarios. Qualitatively 
there are two competing effects that need to be considered. 
The opening up of a pseudogap in the one-particle excitation
spectrum implies that one-particle contribution to the d.c. resistivity
$\rho$ will show an insulating upturn: increasing $\rho$ as $T$ decreases.
However, we must also include the collective contribution to transport
from the pairs, which will show up as a ``Drude-like'' peak in the 
the conductivity (a broadened version of the SC state delta function).
This contribution, by itself, will show a marked down-turn in the
d.c. resistivity with decreasing temperature (paraconductivity).
The net effect of both these contributions need to be considered,
and it is clear that sufficiently close to $T_c$ the pair-channel
will dominate.
(Interestingly, a similar interplay between the depleted one-particle
contribution and the enhanced pair contribution was seen in the
QMC calculation of the compressibility in Sec.~V).

As described in Sec II, experimentally the d.c. resistivity shows
only a distinct, but relatively modest (on the scale of other pseudogap
effects) decrease away from linearity as $T$ decreases.
It is often suggested that the spin-charge separation theories, briefly
discussed in Section III, are able to naturally account for this. However,
there are no quantitative calculations of charge transport by
bosons interacting with gauge fields, and the arguments within that
framework are also, at the present time, qualitative.

A second important point about transport in the real materials
is that it is dominated by regions of ${\bf k}$-space near the 
zone diagonal where the electronic states are much more dispersive
(large Fermi velocity) than those near $(\pi,0)$; 
see [\onlinecite{MR_JC}] and references therein.
This is also where the pseudogap effects are smallest.
These strong ${\bf k}$-dependences of the the electronic structure,
the gaps and the interactions are likely to be crucial for a
proper understanding of transport (and other) experiments.
In particular, the opening up of the pseudogap
starting near the $(\pi,0)$ point at $T^*$, 
and the way in which it progressively affects larger regions
of the Fermi surface with decreasing $T$, as described in Section VIII, 
will be an important input to detailed transport calculations,
as well for understanding the crossovers in other correlation functions.

The effects of magnetic fields in the pseudogap regime,
diamagnetism, and magnetotransport are all questions for
further study. The disconnected Fermi arcs of Section VIII
may have profound consequences in the presence of magnetic fields.

Much remains to be understood about c-axis transport.
As far as the pseudogap in c-axis optical conductivity is concerned
one might ask why there are no low lying charge fluctuations like
those seen in a-b plane transport? The qualitative reason is 
that tunneling of incoherent pairs along the c-axis 
is exponentially suppressed relative to the
one-particle contribution (which is itself small). Hence transport 
across planes is dominated by the one-particle contribution which upon 
the opening of a pseudogap leads to an upturn in the c-axis resistivity
with decreasing temperature.

Another very important question which I have not touched on in 
these lectures is: why are the quasiparticles as seen by
ARPES destroyed above $T_c$? In the pseudogap regime, there is
a suppression of low-lying particle-hole excitations, but nevertheless
an electron can scatter off the pairs. As soon as the pairs become 
phase coherent below $T_c$ this scattering mechanism freezes out.
Detailed calculations along these lines are needed to see if
this mechanism is able to account for the experimental observations.

There is an important question on which more
experimental input would be welcome. There seems to be some
difference in the behaviour of the underdoped bi-layer materials
compared with those which have a single CuO plane per unit cell
(in particular the 214 LaSrCuO). There are suggestions that the
spin-gap is entirely due to bilayer effects [\onlinecite{MM,PWA}].
The recent ARPES observation of a pseudogap in the one-layer
Bi-2201 [\onlinecite{HARRIS_97}] casts doubt on whether bilayers
are necessary for this phenomenon. All of the models described
in these lectures are single-plane models. Nevertheless, more studies of
one-layer cuprates may shed light on whether or not there is an
important distinction between one and two layer systems, especially in so far
as the pseudogap is concerned.

Finally. direct experimental probes of pairing correlations above $T_c$
would be very important. Possible avenues for research include
characteristic signatures in low frequency collective transport, 
and precursors of the Josephson effect between a pseudogap metal 
and a superconductor [\onlinecite{DISCUSS}].


\medskip
\noindent
{\bf X. Conclusions}
\medskip

The main message of these lectures is that, in a sense, the pseudogap
anomalies are {\it not} very mysterious. Although such deviations
from Fermi liquid behaviour above $T_c$ are without precedent,
nevertheless, on theoretical grounds they follow as soon as one
has a superconducting ground state at $T=0$ with a small superfluid
stiffness and a large SC energy gap. In the simple models discussed 
above this was accomplished by being in a regime of low density and
small pair size. Many open questions remain, including some questions
of principle, as discussed in the preceding Section. 
Nevertheless, the pseudogap anomalies, by themselves, do not appear
to require consideration of novel ground states or new types of
excitations. Whether the non-Fermi-liquid anomalies above $T^*$,
and in particular, those at optimal doping, (which were not discussed
in this lecture) force us to consider such new states, or these
are associated with even more complicated
finite temperature crossovers, is not clear at the moment.

The key open questions are: why do the underdoped
cuprates have such a small superfluid density -- which is surely
related to their being doped Mott insulators -- and such a large
effective pairing interaction?
Quantitative comparison with the experiments must, therefore, 
await a controlled calculation based on a microscopic model
which describes how a Mott insulator upon doping
goes into a short coherence-length d-wave superconductor whose
normal state is dominated by pairing correlations.


\medskip
\noindent{\bf Acknowledgements}

I would like to thank Juan-Carlos Campuzano,
Elbio Dagotto, Hong Ding, Jan Engelbrecht, Sasha Nazarenko, Mike Norman, 
Nandini Trivedi, and Andrei Varlamov for very fruitful collaborations.
In addition I acknowledge useful conversations with
B. Battlogg, \O. Fischer, W. Y. Liang, J. Loram, 
A. Paramekanti, T. V. Ramakrishnan, B. S. Shastry,
Z. X. Shen, T. Timusk, and Y. J. Uemura.
Finally, I would like to thank Professor J. R. Schrieffer and
Professor G. Iadonisi for inviting me to Varenna, and 
Dr.~Maria Luisa Chiofalo and all the participants for making
it a very enjoyable and stimulating summer school.


\begin{figure}
\vspace{50pt}
  {\epsfxsize=5.0truein\epsfbox{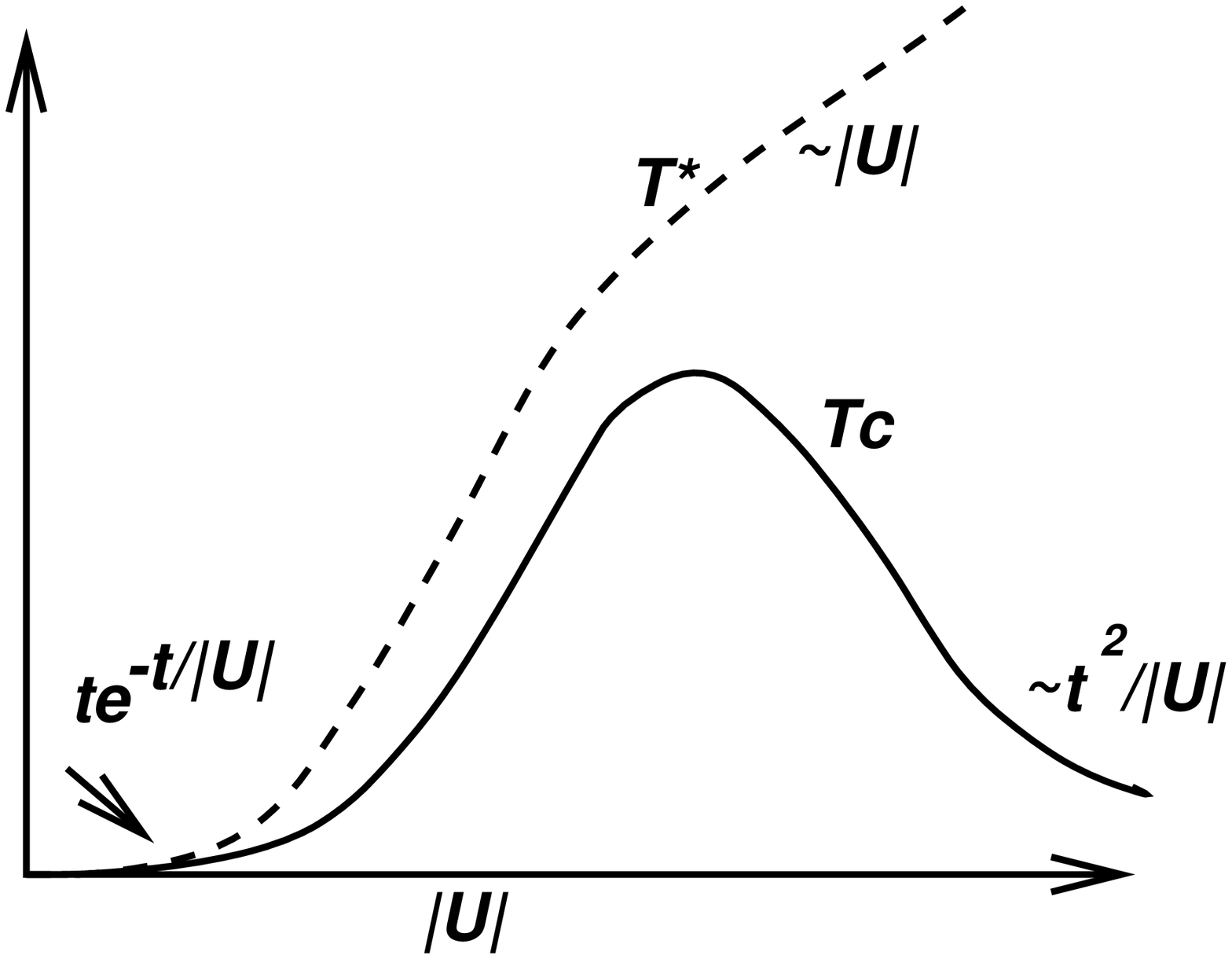}}
\vskip 0.5truecm
\caption{
Schematic phase diagram of the attractive Hubbard model illustrating the
crossover from a BCS superconductor with a Fermi liquid normal
state (in the weak coupling $|U|/t \ll 1$ limit) to a condensate of bosons
with a normal Bose liquid state above $T_c$ (for $|U|/t \gg 1$). 
$T_c$ is the transition temperature at which
phase coherence is established and $T^\ast$ is a crossover scale
below which pairing correlations become manifest in the normal state.
\label{1}}
\end{figure}

\vfill\eject

\begin{figure}
\vspace{0pt}
  {\epsfxsize=3.5truein\epsfbox{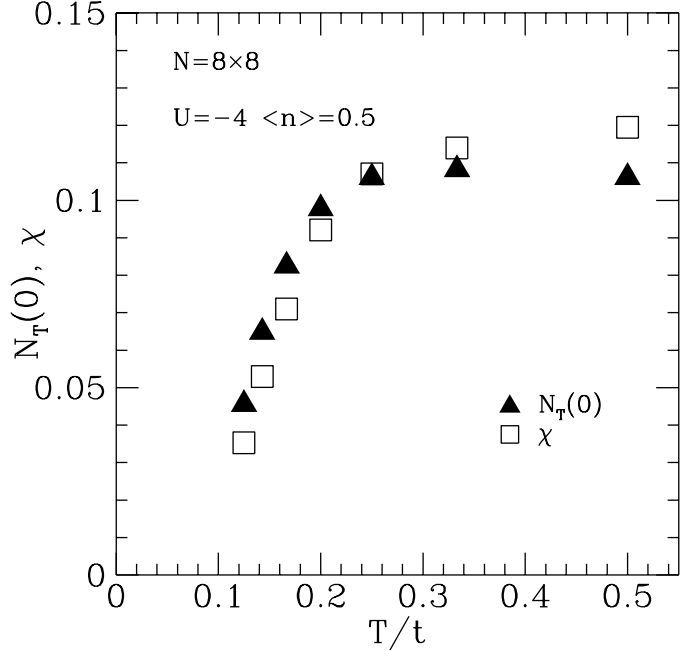}}
\vskip 0.5truecm
\caption{
The one-particle density of states $N_T(0)$ at the chemical potential
(full triangles) and the spin susceptibility $\chi_s(T)$ (open squares)
as a function of temperature. These results were obtained from Quantum
Monte Carlo simulations [15,17] of the 2D attractive
Hubbard model for $|U|/t = 4$, $\langle n \rangle = 0.5$ and $L^2 = 8^2$.
}
\label{2}
\end{figure}

\begin{figure}
\vspace{0pt}
  {\epsfxsize=3.5truein\epsfbox{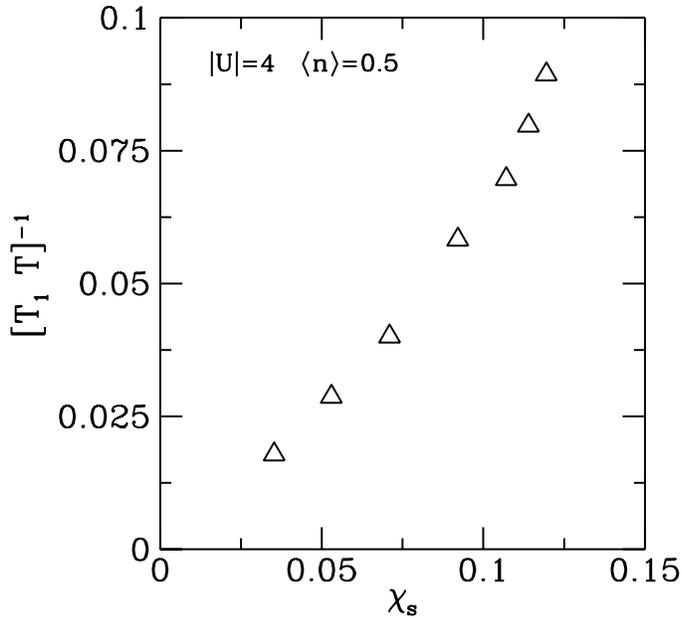}}
\vskip 0.5truecm
\caption{
Parametric plot of NMR relaxation rate $1/T_1T$ vs
spin susceptibility $\chi_s$ of the 2D attractive Hubbard model
for $|U|/t = 4$, $\langle n \rangle = 0.5$ with temperature as the
implicit variable.
(Adapted from ref.~[15]).
The plot shows that that the two quantities track each other:
$1/T_1T \sim \chi_s(T)$. Note that for a Fermi liquid all the data
would collapse to a single point.
}
\label{3}
\end{figure}

\vfill\eject

\begin{figure}
\vspace{0pt}
  {\epsfxsize=3.5truein\epsfbox{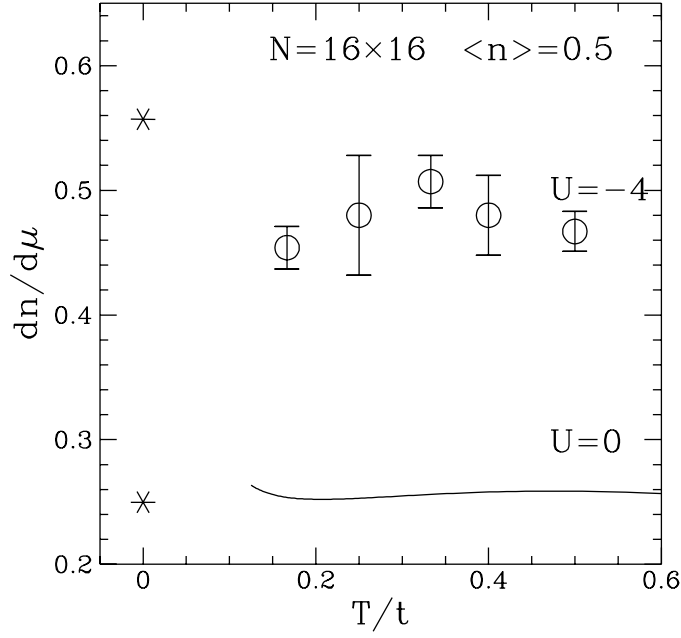}}
\vskip 0.5truecm
\caption{
The compressibility $dn/d\mu$ for $\langle n\rangle = 0.5$, $U = 0$ (full line)
and $|U| = 4$ (open circles), as a function of $T$ obtained on a
$16^2$ lattice.  The $T=0$ non-interacting and the $T=0$ mean field result
for $|U| = 4$ are also shown. (From ref.~[17]).
}
\label{4}
\end{figure}

\begin{figure}
\vspace{0pt}
  {\epsfxsize=3.5truein\epsfbox{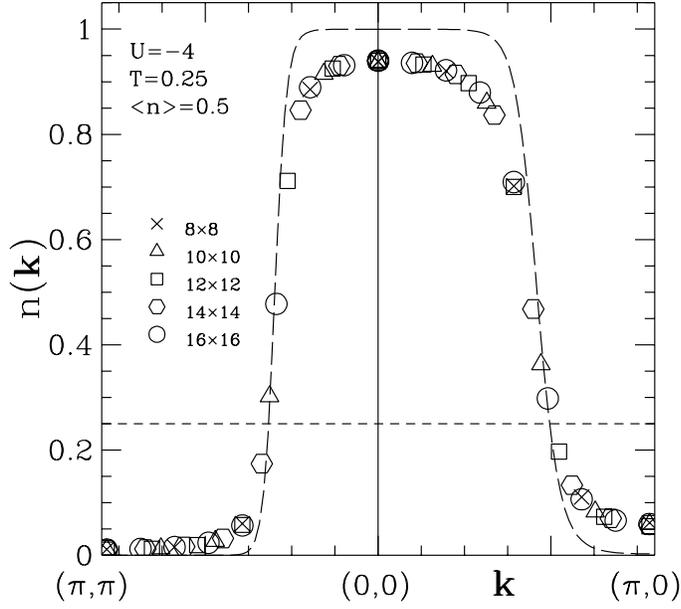}}
\vskip 0.5truecm
\caption{
The momentum distribution function $n({\bf k})$
as a function of ${\bf k}$ for $|U| = 4t$, $\langle n \rangle = 0.5$,
and $T = 0.25t$. (From ref.~[17]).
}
\label{5}
\end{figure}

\vfill\eject

\begin{figure}
\vspace{0pt}
  {\epsfxsize=3.5truein\epsfbox{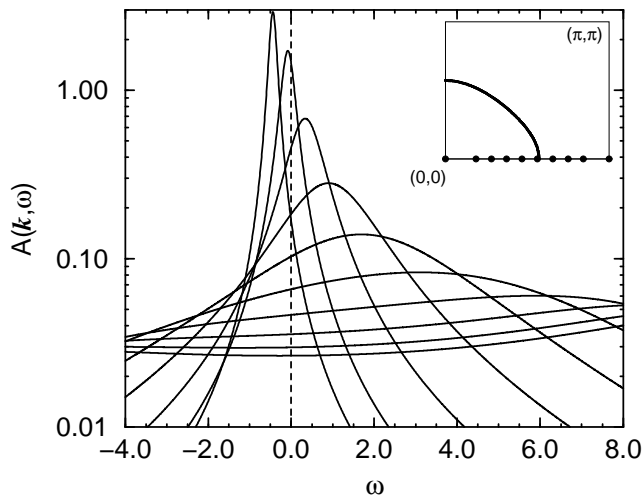}}
\vskip 0.5truecm
\caption{
Self-consistent $T$-matrix results for the spectral function
$A({\bf k},\omega)$ for a sequence of momenta
${\bf k}=(x\pi/32,0)$ with $x=\{0,6,9,12,15,18,21,24,27,32\}$
for $U_d=-8$ and $n=0.5$ at a high temperature $T=2.0$ ($T > T^*\simeq 1$).
Inset: Brillouin zone with points indicating the momenta ${\bf k}$ and
solid curve showing the $T=0$ non-interacting Fermi surface.
(From ref.~[42]).
}
\label{6}
\end{figure}

\begin{figure}
\vspace{0pt}
  {\epsfxsize=3.5truein\epsfbox{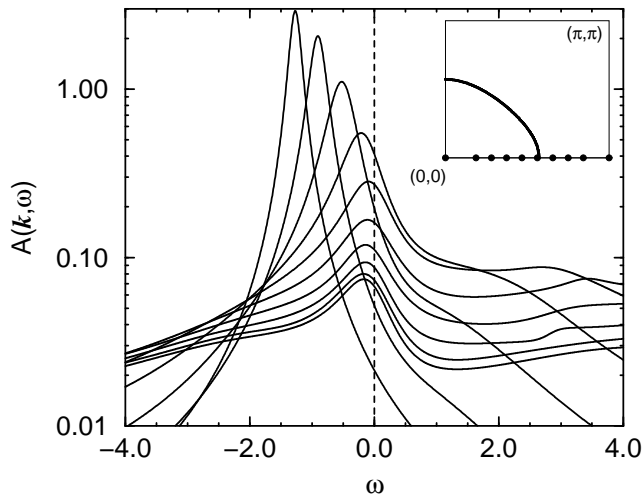}}
\vskip 0.5truecm
\caption{
Spectral functions $A({\bf k},\omega)$ for the same momenta
as in Fig.~6: ${\bf k}=(x\pi/32,0)$ with $x=\{0,6,9,12,15,18,21,24,27,32\}$
for $U_d=-8$ and $n=0.5$, but at a temperature $T=0.2$ below $T^* \simeq 1$.
Notice the changes in lineshape and anomalous dispersion compared with
high temperature results of Fig.~6.
(From ref.~[42]).
}
\label{7}
\end{figure}

\vfill\eject

\begin{figure}
\vspace{0pt}
  {\epsfxsize=3.5truein\epsfbox{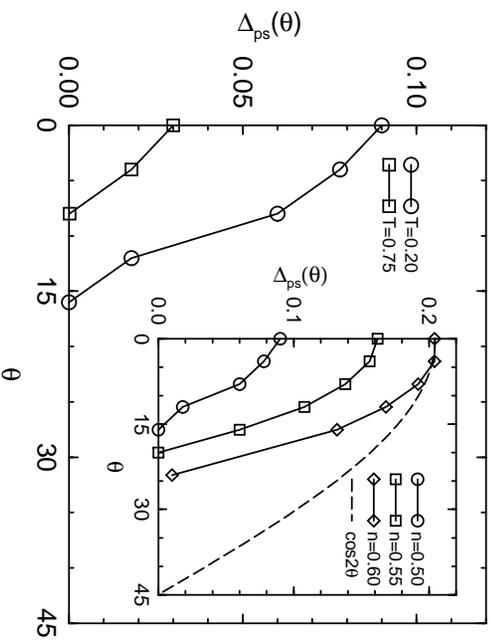}}
\vskip 0.5truecm
\caption{
The angle dependence of the pseudogap $\Delta_{ps}(\theta)$
for temperatures $T=0.2$ and  $T=0.75$ (both below $T^*$)
and $U_d=-8$, $n=0.5$.
Inset: The angle dependence of $\Delta_{ps}(\theta)$
for various densities $n=0.5$, $n=0.55$  and  $n=0.60$
and $U_d=-8$, $T=0.2$.
(From ref.~[42]).
}
\label{8}
\end{figure}

\begin{figure}
\vspace{0pt}
  {\epsfxsize=3.5truein\epsfbox{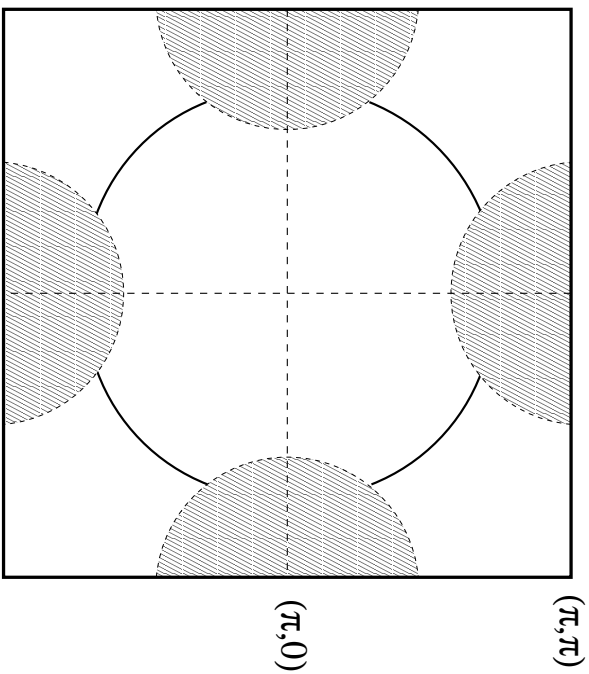}}
\vskip 0.5truecm
\caption{
Schematic representation of the results of ref.~[42]
showing that the Fermi surface is destroyed in patches.
Solid lines represent gapless excitations and the shaded regions
indicate momenta at which the pseudogap opens up and
the Fermi surface is destroyed.
}
\label{9}
\end{figure}


\end{document}